\documentclass[10pt]{article}

\begin{document}

\title{Exterior spacetime for stellar models in $5$-dimensional  Kaluza-Klein gravity}
\author{J. Ponce de Leon\thanks{E-mail: jpdel@ltp.upr.clu.edu, jpdel1@hotmail.com}\\ Laboratory of Theoretical Physics, Department of Physics\\ 
University of Puerto Rico, P.O. Box 23343, San Juan, \\ PR 00931, USA} 
\date{January  2007}

\maketitle

\begin{abstract}

It is well-known that Birkhoff's theorem is no longer valid in theories with more than four dimensions.
Thus,  in these theories the {\it effective} four-dimensional picture allows the existence   
of  different possible, non-Schwarzschild, scenarios for the description of the spacetime outside of a spherical star, contrary to four-dimensional general relativity. We investigate the exterior spacetime of a spherically symmetric star in the context of Kaluza-Klein gravity. We take a well-known family of static spherically symmetric solutions of the Einstein equations in an {\it empty} five-dimensional universe, and analyze possible stellar exteriors that are conformal to the metric induced on four-dimensional hypersurfaces orthogonal to the extra dimension. We show that all these non-Schwarzschild exteriors can continuously be matched with the interior of the star, indicating that the matching  conditions at the boundary of a star do not require an unique exterior. Then, without making any assumptions about the interior solution, we prove the following statement: the condition that in the weak-field limit we recover the usual Newtonian physics singles out an unique exterior. This exterior is ``similar" to Scharzschild vacuum in the sense that it has no effect on gravitational interactions. However, it is more realistic because instead of being absolutely empty, it is consistent with the existence of quantum zero-point fields {(P.S. Wesson, {\em Phys. Essays, Orion} {\bf 5}, 591(1992))}.
 We also examine the question of how would the deviation from the Schwarzschild vacuum exterior affect the parameters of a neutron star. In the context of a model star of uniform density, we show that the general relativity upper  limit $M/R < 4/9$ is significantly {\it increased} as we go away from the Schwarzschild vacuum exterior. We find that, in principle, the compactness limit of a star can be larger than $1/2$, without being a black hole. The generality of our approach is also discussed.

\end{abstract}

\medskip

PACS: 04.50.+h; 04.20.Cv

{\em Keywords:} Kaluza-Klein Theory; Space-Time-Matter Theory; General Relativity. 

\newpage
\section{Introduction}

In Newtonian and relativistic theories of gravity the simplest models of isolated stars are provided by static spherically symmetric distributions of matter surrounded by empty space. 
In general relativity, the  interior of a star is modeled by a  solution of the  Einstein field equations with some energy-momentum tensor satisfying physical conditions. 
The exterior of an isolated star is described by the Schwarzschild metric, which according to Birkhoff's theorem is the unique asymptotically flat vacuum solution with spatial spherical symmetry.  
At the boundary of the star, which is a spherical surface of zero-pressure, both solutions have to be  matched following a standard procedure in general relativity. These models are used to study the physical features of compact objects such as white dwarfs and  neutron stars.

Today, there are a  number of theories that envision our four-dimensional world  as embedded in a larger  universe with more than four dimensions.  In these theories it is natural to look for stellar models, similar to those of general relativity,  and study the consequences of the existence of extra dimensions on the structure of neutron stars. 
However, there is an important aspect that complicates the construction of such models in these theories. Namely that Birkhoff's theorem is no longer valid in more than four dimensions, i.e., there is no an unique asymptotically flat vacuum solution with spatial spherical symmetry. As a consequence, the {\it effective} theory in $4D$ allows the existence   
of  different possible, non-Schwarzschild, scenarios for the description of the spacetime outside of a spherical star, contrary to four-dimensional general relativity. 

In braneworld models, Germani and Maartens \cite{Cristiano} have found  two exact vacuum solutions, both asymptotically Schwarzschild, which  can be used to represent the  exterior of an uniform-density star on the brane. The analysis of the gravitational  collapse and black hole formation on the brane is even more complicated; Bruni, Germani and Maartens \cite{Bruni} have shown that the vacuum exterior of a collapsing dust cloud {\it cannot} be static, although, on general physical grounds it is expected that the non-static behavior will be transient, so that the exterior tends to a static form.  Similar results, for the collapse of a dust cloud, have been derived by Kofinas and Papantonopoulos \cite{Kofinas} in the context of various braneworld models with curvature corrections. 

However, in Kaluza-Klein theories a  milder version of Birkhoff's theorem is true. Namely,  
there is only {\it one} family of  spherically symmetric exact solutions of the field equations $R_{AB} = 0$ which are asymptotically flat, static and independent of the ``extra" coordinates. In five-dimensions, in the form given by Davidson and Owen \cite{Davidson Owen}, it is described by the line element
 \begin{equation}
\label{Davidson and Owen solution}
dS^2 = \left(\frac{ar - 1}{ar + 1}\right)^{2\sigma k}dt^2 - \frac{1}{a^4r^4}\frac{(ar + 1)^{2[\sigma(k - 1) + 1]}}{(ar - 1)^{2[\sigma(k - 1) - 1]}}[dr^2 + r^2d\Omega^2]
\pm  \left(\frac{ar + 1}{ar -1}\right)^{2\sigma}dy^2,
\end{equation}
where $a$ is a constant with dimensions of $L^{- 1}$; and  $\sigma$ as well as  $k$ are parameters that obey the constraint 
\begin{equation}
\label{constraint on sigma and k}
\sigma^2(k^2 - k + 1) = 1.
\end{equation} 
This family of solutions has been rediscovered in different forms by Kramer \cite{Kramer} and, although in a different context,  by  Gross and Perry \cite{Gross Perry} (for  a recent discussion see \cite {JPdeLgr-qc/0611082} and references therein). Its importance resides in the fact that in the limit  $k \rightarrow \infty$ $(\sigma k \rightarrow 1)$, it  yields\footnote{This metric is known as the Schwarzschild black string.} 
\begin{equation}
\label{Schw. limit}
dS^2 = \left(\frac{1 - 1/ar}{1 + 1/ar}\right)^2dt^2 - \left(1 + \frac{1}{ar}\right)^4[dr^2 + r^2d\Omega^2]
\pm  dy^2,
\end{equation}
which on every subspace $y$ = constant reduces to the Schwarzschild metric, in isotropic coordinates, for a central mass $M = 2/a$. Namely,  
\begin{equation}
\label{Schw. in isotropic coordinates}
ds^2 = \left(\frac{1 - M/2r}{1 + M/2r}\right)^2dt^2 - \left(1 + \frac{M}{2r}\right)^4[dr^2 + r^2d\Omega^2].
\end{equation}

 In  five-dimensional  Kaluza-Klein theory, these solutions play a central role in the discussion of  
many important observational problems, which include the classical tests of relativity, as well as the geodesic precession of a gyroscope and possible departures from the equivalence principle \cite{Wesson book}. In the context of the induced-matter approach, the metric (\ref{Davidson and Owen solution}) is interpreted as describing extended spherical objects called solitons \cite{Wesson-JPdeL}. In this interpretation the matter distribution contains a lightlike singularity at $r = 1/a$ \cite{JPdeLgr-qc/0611082} \cite{Lake}, which, in principle, can be visible to an external observer. Only in the black string limit  $k \rightarrow \infty$ $(\sigma k \rightarrow 1)$ the $5D$ metric (\ref{Davidson and Owen solution}) possesses an event horizon; for any finite value of $k$ the horizon is reduced to a singular point.

However, the presence of naked singularities makes everybody uncomfortable. 
In order to avoid such singularities, we have to exclude the central region and require $r > 1/a$. What this suggests is that the asymptotically flat metric (\ref{Davidson and Owen solution}) should be interpreted   as an ``exterior" 
solution, describing the gravitational field outside of the core of a spherical matter distribution. In this interpretation, the effective exterior is {\it not} empty because there are  nonlocal stresses induced from the Weyl curvature in $5D$, which in $4D$ behave like radiation.\footnote{It is sometimes called ``black" or ``Weyl" radiation, because in this case the induced energy-momentum tensor $T_{\mu\nu}$ can be expressed as $T_{\mu\nu} = - \epsilon E_{\mu\nu}$, where $E_{\mu\nu}$ represents the spacetime projection of the five-dimensional Weyl tensor, which is traceless.}
  The ``interior" region $r \leq 1/a$  has to  be described by some another solution of the field equations, which must be regular at the origin and  not necessarily asymptotically flat. The aim of this work is to investigate  this interpretation and its consequences on stellar models. 

In this context, the simplest candidate to describe the exterior of a spherical star is given by the spacetime part of the $5D$ vacuum solution (\ref{Davidson and Owen solution}), which is the metric induced on every hypersurface $y$ = constant. 
Nevertheless, we show here that this simple exterior leads to stellar models  that have no Newtonian limit. On the other hand, according to the {\it correspondence principle}, we should require that our stellar models be consistent not only with general relativity but also with Newtonian models. Therefore,  in this work we consider a more general  scenario. The paper is organized as follows.

In section $2$ we use the five-dimensional line element (\ref{Davidson and Owen solution}) to generate a family of asymptotically flat metrics in $4D$ which are conformal to the metric induced on $y$ = constant hypersurfaces. 
 We examine this family as a possible choice for describing  the  four-dimensional spacetime outside of a spherical star of radius $r_{b} > 1/a$. 

In section $3$, from the boundary conditions, we find that there is only one member of this family compatible with the weak-field limit. In other words, the requirement that in the weak-field limit we recover the usual Newtonian physics singles out an unique exterior. This exterior depends on one parameter, which we call $\varepsilon$, and includes the vacuum Schwarzschild exterior for $\varepsilon = 1$. The intriguing feature of this metric is that it is ``similar" to the Scharzschild vacuum in the sense that it has no effect on gravitational interactions. However, it is more realistic because instead of being absolutely empty, it is consistent with the existence of quantum zero-point fields \cite{WessonEssay}.

In section $4$, in order to illustrate our results we examine in detail the particular model where the central core is represented by a  perfect fluid sphere of homogeneous incompressible fluid. We calculate the physical parameters of a neutron star for distinct  exteriors and different degrees of deviation from the Schwarzschild metric. We show that the general relativity compactness limit $M/R < 4/9$ is significantly increased as we go away from the Schwarzschild vacuum exterior. In fact,  we find that in principle $M/R$ can be larger than $1/2$, without being a black hole. 
Finally, in section $5$ we give a summary of our work and discuss the generality of our results.

\section{The model}

In this section we present the main steps of our work. First,  we address the question  of how to identify the effective four-dimensional spacetime. 
The correct answer to this question is crucial in order to be able to predict effects from an extra dimension. Then,  from (\ref{Davidson and Owen solution}),  we construct the ``appropriate" set of possible exterior metrics.  Next, we describe the model for the central core and examine the  
boundary conditions, which ensure that the interior and exterior metric join continuously. 

\subsection{The effective spacetime}

In five-dimensional  models, the effective four-dimensional picture\footnote{This is the picture registered by an observer who is not aware of the existence of an extra dimension.} is usually recovered, or constructed, from $g_{\mu\nu}^{ind}$ the four-dimensional metric induced on $4D$ hypersurfaces that are orthogonal to  what is considered the ``extra" dimension \cite{JPdeL:gr-qc/0512067}. For a given line element in $5D$  
\begin{equation}
\label{general metric}
dS^2 = g_{\mu\nu}(x^{\rho}, y)dx^{\mu}dx^{\nu} + \epsilon \Phi^2(x^{\rho}, y)dy^2,\;\;\;\epsilon = \pm 1,
\end{equation}
the induced metric, on any hypersurface  $y =$ constant, is just $g_{\mu\nu}$, i.e., $g_{\mu\nu}^{ind} = g_{\mu\nu}$. The question is how to obtain our physical spacetime, with metric $g_{\mu \nu}^{eff}$, from  the induced metric $g_{\mu\nu}^{ind}$. 

Unfortunately, there is no a consensus answer to this question. In fact,  there are different competing approaches which are not equivalent. Today, in braneworld models, as well as in space-time-matter (STM) theory, the standard technique is to identify the metric of our spacetime with the induced one, i.e., $g_{\mu\nu}^{eff} = g_{\mu\nu}^{ind}$. However, there are a number of theories and approaches where $g_{\mu\nu}^{eff}$ is constructed in a different ways.  

The simplest example is given by a metric which  has been termed  canonical \cite{Mashhoon}, where the $g_{\mu\nu}^{eff}$ is conformal to $g_{\mu\nu}^{ind}$, with a conformal factor depending only on the extra coordinate. Also, 
a number of  authors \cite{Wesson book} \cite{Kokarev1} \cite{Kokarev2} \cite{Sajko} \cite{JPdeL:gr-qc/0105120} \cite{Horowitz}  have considered models where $g_{\mu\nu}^{eff}$ is conformal to $g_{\mu\nu}^{ind}$, with a conformal factor that is a ``function" of the scalar field $\Phi$. 

\medskip

In this work we examine  the consequences of $4D$ models obtained from a dimensional reduction of the metric in $5D$, on a hypersurface $y$ = constant, with a conformal rescaling 
\begin{equation}
\label{general factorization}
g^{eff}_{\mu \nu} = \Phi^N g_{\mu\nu}^{ind},
\end{equation}
where $N$ is an arbitrary real constant. This {\it ansatz} consolidates various approaches in the literature. In the context of Kaluza-Klein gravity and STM, it has been considered by Wesson \cite{Wesson book}, Kokarev \cite{Kokarev1}-\cite{Kokarev2}, Sajko \cite{Sajko}, and the present author \cite{JPdeL:gr-qc/0105120}. 
For $N = 0$, it reproduces the usual interpretation of  braneworld and STM theories. For $N = -2$ the interpretation is similar to the one provided by the canonical metric. 
For $N = 1$ it yields  the classical Davidson-Owen \cite{Davidson Owen} and Dolan-Duff \cite{Dolan} interpretation.

\subsection{Possible non-Schwarzschild stellar exteriors }

 The effective four-dimensional spacetime (\ref{general factorization}), corresponding to the {\it empty} five-dimensional universe described by metric  (\ref{Davidson and Owen solution}),  can be written as 
\begin{equation}
\label{eff. spacetime for Davidson and Owen solution}
ds^2 = \left(\frac{ar - 1}{ar + 1}\right)^{2\varepsilon}dt^2 - \frac{(ar + 1)^2(ar - 1)^2}{a^4r^4}\left(\frac{ar + 1}{ar - 1}\right)^{2[\varepsilon + \sigma(N - 1)]}[dr^2 + r^2d\Omega^2],
\end{equation}
where $\varepsilon = \sigma(k - N/2)$ and\footnote{We note that for $N = 1$ the choice of $\sigma$ is immaterial, because the metric depends only on $\varepsilon$. However, we have selected a positive $\sigma$ in (\ref{sigma in terms of varepsilon})  for consistency with our previous work \cite{JPdeLgr-qc/0611082}.} 
\begin{eqnarray}
\label{sigma in terms of varepsilon}
\sigma &=& \frac{2\left[\varepsilon (1 - N) + \sqrt{(N^2 - 2N + 4) - 3\varepsilon^2}\right]}{N^2 - 2N + 4}, \;\;\;\mbox{for}\;\;\; N 
\geq 1;\nonumber \\
\sigma &=& \frac{2\left[\varepsilon (1 - N) - \sqrt{(N^2 - 2N + 4) - 3\varepsilon^2}\right]}{N^2 - 2N + 4}, \;\;\;\mbox{for}\;\;\; N < 1.
\end{eqnarray}
This parameterization is useful because the ``deviation" from the exterior Schwarzschild metric can be measured by $|\varepsilon - 1|$. Since the exterior spacetime is not in general empty, from a physical point of view taking $\varepsilon \neq 1$ is equivalent to giving the star an  ``atmosphere" extending all the way to infinity with a matter density decreasing as $\rho \sim 1/a^2r^4$. The special case where the star has a sharp surface corresponds to $\varepsilon = 1$, where we recover the exterior Schwarzschild metric, for any value of $N$. In Appendix $A$ we present the equation of state and gravitational mass, of such an atmosphere.
Thus, 
the questions  of interest here are the following:
\begin{enumerate}
\item How does the existence of stars restrict the possible values of $N$?
\item How does a possible deviation from the Schwarzschild vacuum exterior, i.e., going away from $\varepsilon = 1$, can affect the star parameters?

\end{enumerate}
 
The total gravitational mass, measured at infinity, is given by 
\begin{equation}
\label{Total grav. mass}
M  = \frac{2\varepsilon}{a}. 
\end{equation}
Thus, in the Schwarzschild limit $M = 2/a$. Therefore, to ensure the positivity of $M$, we will choose $a > 0$ and  $\varepsilon > 0$ everywhere.

We note that $\sigma(N - 1)$ is invariant under the transformation $N \rightarrow (2 - N)$. This means that  the effective spacetime for $N = 0$ is identical to the one for $N = 2$. The same for $N = 3$ and $N = -1$; $N = 4$ and $N = -2$; and so on\footnote{These are just examples, we are {\it not} assuming that $N$ is an integer number}. In other words: the effective spacetimes  with $N > 1$ duplicate the ones with $N < 1$. Only the one with $N = 1$ remains the same.

\subsection{The interior solution and matching conditions}

In this work we restrict our discussion to the case where the interior solution is time independent. The matching conditions  are easier to derive if we take the line element inside the fluid sphere with the same symmetry properties as the exterior metric (\ref{eff. spacetime for Davidson and Owen solution}). Namely, the interior static  
will be taken as 
\begin{equation}
\label{effective interior metric}
ds^2 = e^{\nu(r)}dt^2 - e^{\lambda(r)}[dr^2 + r^2(d\theta^2 + \sin^2\theta d\phi^2)].
\end{equation}
The three-dimensional surface $\Sigma$, defined by the equation
\begin{equation}
r - r_{b} = 0,
\end{equation}
where $r_{b}$ is a constant, separates the spacetime into two regions: $r \geq r_{b}$ and $r\leq r_{b}$, described by the metrics (\ref{eff. spacetime for Davidson and Owen solution}) and (\ref{effective interior metric}), respectively.

The problem of joining two distinct metrics across surfaces of discontinuity has been discussed by a number of authors (see \cite{Robson} and references therein). Robson \cite{Robson} showed the complete equivalence of the junction conditions proposed by Lichnerowicz and by O'Brien and Synge at non-null hypersurfaces. The latter are analogous to the requirement that the first and second fundamental forms be continuous at $\Sigma$. 

For the line elements under consideration, the first fundamental form at $\Sigma$, becomes 
\begin{equation}
\label{continuity of ds}
ds^2_{(\Sigma: r = r_{b})} = e^{\nu(r_{b})}dt^2 - e^{\lambda(r_{b})} r_{b}^2(d\theta^2 + \sin^2\theta d\phi^2)].
\end{equation}
Therefore, its continuity across $\Sigma$ requires
\begin{eqnarray}
e^{\nu(r_{b})} &=& \left(\frac{ar_{b} - 1}{ar_{b} + 1}\right)^{2\varepsilon}, \nonumber \\
e^{\lambda(r_{b})} &=& \frac{(ar_{b} + 1)^2(ar_{b} - 1)^2}{a^4r_{b}^4}\left(\frac{ar_{b} + 1}{ar_{b} - 1}\right)^{2[\varepsilon + \sigma(N - 1)]}.
\end{eqnarray}
Besides, the ``physical'' radius $R_{b}$ of the sphere with coordinate radius $r_{b}$ is given by 
\begin{equation}
\label{phys. rad. from bound. Cond.}
R_{b} = r_{b}e^{\lambda(r_{b})/2} = \frac{(a^2r_{b}^2 - 1)}{a^2 r_{b}}\left(\frac{ar_{b} + 1}{ar_{b} - 1}\right)^{[\varepsilon + \sigma(N - 1)]}.
\end{equation}
The second fundamental form, say $dK^2$, at $\Sigma$ is given by 
\begin{equation}
dK^2_{\Sigma} = n_{\mu;\nu}dx^{\mu}dx^{\nu},
\end{equation}
where $n^{\mu}$ is the unit vector orthogonal to $\Sigma$. In the present case, if we take
\begin{equation}
n_{\mu} = \delta_{\mu}^{1}e^{\lambda/2}, 
\end{equation}
then we find
\begin{equation}
dK^2_{(\Sigma: r = r_{b})} = - e^{- \lambda(r_{b})/2}\left[\frac{\nu'(r_{b})}{2}\left(e^{\nu(r_{b})/2}dt\right)^2 - \left(\frac{1}{r_{b}} + \frac{\lambda'(r_{b})}{2} \right)\; r_{b}^2 e^{\lambda(r_{b})}\;\left(d\theta^2 + \sin^2\theta\; d\phi^2\right)\right].
\end{equation}
Consequently, in isotropic coordinates, the requirement of continuity of the second fundamental form across $\Sigma$ is equivalent to demand  that the first derivatives of the metric be continuous across\footnote{Thus, in isotropic coordinates the ``weaker" junction conditions of O'Brien and Synge are equivalent to those of  Lichnerowicz.} $\Sigma$.

For the exterior metric (\ref{eff. spacetime for Davidson and Owen solution}), the continuity of $\nu'$ and $\lambda'$ yields 
\begin{eqnarray}
\label{Continuity of the second fund. form}
\nu'(r_{b})&=& \frac{ 4 a\varepsilon}{(a^2 r_{b}^2 - 1)},\nonumber \\
\lambda'(r_{b}) &=&  \frac{4\left\{ 1- [\varepsilon + \sigma(N - 1)]\;ar_{b} \right\}}{r_{b}\;(a^2r_{b}^2 - 1)}.
\end{eqnarray}
From a physical point of view, the continuity of $\nu' $ is equivalent to the continuity of the gravitational mass (\ref{grav. mass in terms of the metric}), in such a way that the force on test particles located at the surface will be the same whether we calculate it from inside or outside.
While, the continuity of $\lambda'$ means that the {\it radial} pressure (i.e. $- T^1_1$) of the fluid at the surface joins continuously with the pressure of space outside the sphere. 

Instead of (\ref{effective interior metric}), we may consider the most general static interior metric, which in curvature coordinates can be written as
\begin{equation}
\label{interior metric in curvature coordinates}
ds^2 = e^{\omega(R)}dt^2 - e^{\mu(R)}dR^2 - R^2(d\theta^2 + \sin^2\theta d\phi^2).
\end{equation}
Now the continuity of the second fundamental form does not require the continuity of $d\mu/dR$. However, by virtue of the equivalence between Lichnerowicz's and O'Brien-Synge's junction conditions, this and the exterior metric match if the mass and radial pressure are continuous at $\Sigma$.

\paragraph{Embedding the interior solution in $5D$:} The interior solution (\ref{effective interior metric}), as well as (\ref{interior metric in curvature coordinates}), can be embedded in the five-dimensional world by foliating the $5$-dimensional manifold as a set of $4$-dimensional hypersurfaces (or slices), in such a way that the effective  metric in $4D$  coincides with the induced one on every  $y =$ constant   hypersurface. 
The embedding can be taken as
 
\begin{equation}
\label{general interior solution}
dS^2 = e^{\nu(r)}dt^2 - e^{\lambda(r)}[dr^2 + r^2(d\theta^2 + \sin^2\theta d\phi^2)] \pm dy^2.
\end{equation}
which physically corresponds to a   star carrying no-scalar charge \cite{Dereli}, \cite{Liddle}.    The corresponding field equations in $5D$ are 
\begin{equation}
G_{AB} = k_{(5)}^2 {^{(5)}T}_{AB},
\end{equation}
where $k_{(5)}^2$ is a constant introduced for dimensional considerations, ${^{(5)}T}_{AB}$ is the energy-momentum tensor in $5D$ and the non-vanishing components of the Einstein tensor $G_{AB}$ are

\begin{eqnarray}
\label{4D field equations}
G_{0}^{0} &=& - e^{ - \lambda}\left(\lambda'' + \frac{\lambda'^2}{4} + 2 \frac{\lambda'}{r}\right), \nonumber \\
G_{1}^{1} &=& - e^{ - \lambda}\left(\frac{\lambda'^2}{4} + \frac{\lambda'\nu'}{2} + \frac{\lambda' + \nu'}{r}\right), \nonumber \\
G_{2}^{2} &=& - e^{ - \lambda}\left(\frac{\lambda'' + \nu ''}{2} + \frac{\nu'^2}{4} + \frac{\lambda' + \nu'}{2r}\right),
\end{eqnarray}
 $G_{3}^{3} = G_{2}^{2}$, and 
\begin{equation}
\label{p5}
G_{4}^{4} = - e^{- \lambda}\left(\lambda'' + \frac{\lambda'^2}{4} + 2\frac{\lambda'}{r} + \frac{\nu' \lambda'}{4} + \frac{\nu'}{r} + \frac{\nu'^2}{4} + \frac{\nu''}{2}\right).
\end{equation}
We note the relationship $G_{0}^{0} + G_{1}^{1} + 2 G_{2}^{2} = 2 G_{4}^{4}$. Thus, taking 
${^{(5)}T}^{A}_{B} = \;$diag$(\rho, - p_{r}, - p_{\perp}, - p_{\perp}, - p_{5})$ (see equation (\ref{components of EMT})) we find that the matter in  $5D$ satisfies the equation of state\footnote{Here $\rho, p_{r}$ and $p_{\perp}$ represent the energy density, radial and tangential ``pressure", respectively; $p_{5}$ is the pressure along the extra dimension.}
\begin{equation}
\label{eq. of state for the core}  
\rho = p_{r} + 2p_{\perp} - 2 p_{5},
\end{equation}
which is the counterpart of the equation of state (\ref{equation of state for N = 0}) for the exterior of a spherical star.
A four-dimensional observer living in a subspace $y$ = constant is not directly aware of (\ref{p5}), the ``pressure"  along the extra dimension, and therefore will not be able to find  such a simple equation of state between the density and pressures. However, this does not mean that she/he loses {\it all} information about the $5$th dimension; it is consolidated in the  nonlocal stresses induced in $4D$ from the Weyl curvature in $5D$.  As a consequence,  even in the absence of matter, the exterior of a spherical star is {\it not} in general an empty Schwarzschild spacetime. We note that in the above discussion the signature of the extra dimension is irrelevant.

An observer in $4D$
interprets the spacetime part of  ${^{(5)}T}_{AB}$ as the energy-momentum in $4D$. More precisely, as 
\begin{equation}
8\pi T_{\alpha \beta} = G_{\alpha\beta}.
\end{equation}
It is clear that the five-dimensional conservation equation ${^{(5)}T}_{A;B}^{B} = 0$, in the present case  yields the  conservation equation $T^{\mu}_{\nu; \mu} = 0$ in $4D$, which is equivalent to the ``generalized" Tolman-Oppenheimer-Volkov equation of hydrostatic equilibrium 
\begin{equation}
\label{generalized TOV equation}
(\rho + p_{r})\frac{M}{r^2}e^{- (\nu + \lambda)/2} = - \frac{dp_{r}}{dr} + \left(\frac{2}{r} + \lambda'\right)({p_{\perp} - p_{r}}).
\end{equation}
Finally, we note that for metrics (\ref{effective interior metric}) and (\ref{general interior solution}) a motion along a geodesic in $4D$ is also geodesic in $5D$. Therefore, a particle located on a hypersurface $y = $ constant, with zero momentum along the extra dimension, will stay on that hypersurface.   

\section{General interior and the weak-field limit}

The main question in this paper is whether the existence of stars leads to some constraints on $N$. In this section we give a positive answer to this question. We use the boundary conditions,  as well as the weak-field limit, to show that the exterior of a spherical star is given by the metric (\ref{eff. spacetime for Davidson and Owen solution}) with $N = 1$; for any other $N \neq 1$  there are problems in the Newtonian limit.

First of all, let us introduce the notation
\begin{equation}
\label{definition of x}
x = ar_{b},
\end{equation}
in terms of which the continuity equations (\ref{Continuity of the second fund. form}) become
\begin{eqnarray}
\label{Continuity of the second fund. form gen}
\frac{\nu'(r_{b})}{4a}&=& \frac{ \varepsilon}{x^2 -1},\;\;\;\;\; x > 1, \nonumber \\
\frac{\lambda'(r_{b})}{4a} &=&  \frac{\left\{ 1- [\varepsilon + \sigma(N - 1)]\;x \right\}}{x\;(x^2 - 1)}.
\end{eqnarray}
These can be used to get an equation for $x$. Namely, 
\begin{equation}
\label{Model independent approach}
x = \left\{{\varepsilon \left[1 + \frac{\lambda'(r_{b})}{\nu'(r_{b})}\right] - \sigma (1 - N)}\right\}^{- 1}.
\end{equation}
In the limit $\varepsilon \rightarrow 1$,  the radial pressure vanishes at the boundary surface, i.e., we recover the usual matching conditions for the Schwarzschild exterior. However, we note that  in the weak-field limit (\ref{Model independent approach}) violates the positivity of $x$. Indeed, in the Newtonian limit (see equation (\ref{f' = - g'}) bellow)
\begin{equation}
\label{lambda'/nu' in the Newtonian limit}
\frac{\lambda'}{\nu'}\sim  - 1.
\end{equation}
On the other hand $\sigma(1 - N) > 0$ for any $N \neq 1$ and $\varepsilon \neq 1$.  Therefore, $\lambda'/\nu'$ {\it cannot} approximate $- 1$ otherwise $x < 0$. This shows that, given the standard matching conditions at the boundary, the effective exterior  of a spherical  star must be described either by the vacuum Schwarzschild spacetime or by  the line element (\ref{eff. spacetime for Davidson and Owen solution}) with $N = 1$, otherwise the model star is inconsistent with a  Newtonian limit. 

\paragraph{Newtonian limit:} In the weak field-limit we set $e^{\nu} = 1 + \xi\;f(r)$ and $e^{\lambda} = 1 + \xi\; h(r)$, where $\xi$ is a ``small" {\it dimensionless} parameter, i.e., $|\xi| \ll 1$. In this limit $\rho \gg p$. Therefore, from (\ref{standard gravitational mass}) it follows that 
\begin{equation}
M(r) = 4\pi \int_{0}^{r}{{\bar{r}}^2 \rho(\bar{r})d\bar{r}} + O(\xi^2).
\end{equation}
In this limit the energy density is given by
\begin{equation}
8\pi \rho = - \xi \left(h'' + \frac{2 h'}{r}\right). 
\end{equation} 
 From which we get $M = - \xi r^2 h'/2$. On the other hand, in this limit, (\ref{grav. mass in terms of the metric}) yields $M = \xi r^2 f'/2$. Consequently,
\begin{equation}
\label{f' = - g'}
f'(r) = - h'(r), \;\;\;\frac{\lambda'}{\nu'} = -1 + \xi (h - f).
\end{equation} 
Then, the pressures are of second order in $\xi$. Namely,
\begin{eqnarray}
8 \pi p_{r} &=& - \frac{\xi^2}{4 r}\left(8 h'h + r h'^2\right) + \frac{\xi^2 h'}{r}(f + h),\nonumber \\
8 \pi p_{\perp} &=& - \frac{\xi^2}{4r}\left(3 rh'^2 + 4 r h h'' + 4 h h'\right) + \frac{\xi^2 (rh'' + h')}{2r}(f + h). 
\end{eqnarray} 
We note that 
\begin{equation}
8\pi\left[ - \frac{dp_{r}}{dr} + \frac{2}{r}(p_{\perp} - p_{r})\right] = \xi^2 \left(\frac{h'^2}{r} + \frac{h' h''}{2}\right).
\end{equation}
The r.h.s. can be written as $-(\xi h'/2)(8 \pi \rho) = 8\pi \rho M/r^2$. Thus, we recover the Newtonian equation of hydrostatic equilibrium, viz., 
\begin{equation}
\frac{M \rho}{r^2} = - \frac{dp_{r}}{dr} + \frac{2}{r}(p_{\perp} - p_{r}),
\end{equation}
which is nothing but the non-relativistic limit of the TOV equation (\ref{generalized TOV equation}).

\paragraph{The non-Schwarzschild stellar exterior:} The conclusion from this section is that, in Kaluza-Klein gravity the only exterior spacetime that is consistent with the existence of a Newtonian limit is the one given by (\ref{eff. spacetime for Davidson and Owen solution}) with $N = 1$. In this case, the exterior of a spherical star takes a particular simple form. Namely, 
\begin{equation}
\label{The only exterior metric compatible with New. Lim}
ds^2 = \left(\frac{ar - 1}{ar  + 1}\right)^{2\varepsilon}dt^2  - \frac{1}{a^4r^4}\frac{(ar + 1)^{2[\varepsilon + 1]}}{(ar - 1)^{2[\varepsilon - 1]}}[dr^2 + r^2d\Omega^2].
\end{equation}
In addition to the equation of state $\rho = - (p_{r} + 2p_{\perp})$ mentioned in (\ref{Eq. of state for N = 1}), we find that the external field obeys a second equation of state, which is 
\begin{equation}
p_{\perp} = - p_{r}.
\end{equation}
In other words; $p_{r} = \rho$ and $p_{\perp} = - \rho$, where the energy density is given by
\begin{equation}
\label{external density for N = 1}
8\pi \rho = \frac{4a^6 r^4(1 - \varepsilon^2)}{(ar + 1)^4(ar - 1)^4}\left(\frac{ar - 1}{ar + 1}\right)^{2\varepsilon},
\end{equation}
and the gravitational mass (\ref{grav. mass}) is constant outside the fluid sphere. Namely,
\begin{equation}
\label{grav. mass for N = 1}
M = \frac{2\varepsilon}{a}.
\end{equation}
This means that the (total) gravitational mass resides entirely inside the fluid sphere; the external cloud has {\it no} influence on the gravitational mass. This is due to the equation of state $\rho = - (p_{r} + 2p_{\perp})$, which describes matter with no effect on gravitational interactions\footnote{This equation of state is the generalization of $\rho = - 3 p$, which has been considered in different contexts by several authors. Notably, in discussions of cosmic strings by Gott and Rees \cite{Gott}, by  Kolb \cite{Kolb}, as well as in certain sources (called ``limiting configurations") for the Reissner-Norstr\"{o}m field  by the present author \cite{Presentauthor}. Besides, it is the only equation of state consistent with the existence of quantum zero-point fields by Wesson \cite{WessonEssay}.} Finally, we note that the conditions   $\rho \geq 0$ and $M > 0$ demand $0 < \varepsilon \leq 1$. For $\varepsilon = 1$ we recover the Schwarzschild exterior.

\section{Model for the stellar interior}
In order to illustrate the above results we need to assume some interior model. Here we consider a model star in the shape of a homogeneous sphere of incompressible fluid with radius $r_{b}$. Thus, inside the sphere the metric will be the interior Schwarzschild metric, and outside $r_{b}$ it will join continuously to the exterior metric (\ref{eff. spacetime for Davidson and Owen solution}).  
Despite their simplicity, uniform-density models give insight about much wider classes of stellar models \cite{Adler}.

The line element describing the interior of a fluid sphere with constant energy density $\rho_{0}$  and isotropic pressures can be written as 
\begin{equation}
ds^2 = A^2\left(\frac{1 + Br^2}{1 + Cr^2}\right)^2dt^2 - \frac{D^2}{(1 + Cr^2)^2}(dr^2 + r^2 d\Omega^2).
\end{equation}
The matter quantities are 
\begin{equation}
8\pi\rho_{0} = \frac{12C}{D^2},\;\;\;C > 0,
\end{equation}
\begin{equation}
8\pi p(r) = \frac{4\left[(B - 2C) + Cr^2(C - 2B)\right]}{D^2(1 + Br^2)}. 
\end{equation}
In order to simplify these expressions we evaluate the pressure and the equation of state at the center. Namely, 
\begin{equation}
8\pi p_{c} = \frac{4(B - 2C)}{D^2},\;\;\;p_{c} = \gamma \rho_{0},
\end{equation}
 where $p_{c} = p(0)$ and $\gamma$ is a constant. 
In terms of these quantities we find: $B = C(2 + 3\gamma)$; $(B - 2C) = 3C\gamma$ and $(C - 2B) = - 3C(1 + 2\gamma)$. Consequently, the appropriate line element and pressure are
\begin{equation}
\label{int. metric}
ds^2 = A^2\left(\frac{1 + (2 + 3\gamma)Cr^2}{1 + Cr^2}\right)^2dt^2 - \frac{3 C}{2\pi \rho_{0}(1 + Cr^2)^2}(dr^2 + r^2 d\Omega^2),
\end{equation}
\begin{equation}
\label{int. pressure}
p = \rho_{0} \frac{[\gamma - (1 + 2\gamma)Cr^2]}{[1 + (2 + 3\gamma)Cr^2]}.
\end{equation}
Since the density and pressure are positive, it follows that
\begin{equation}
\label{positivity of gamma}
\gamma > 0.
\end{equation} 
We note that, in principle, $\gamma$ can take arbitrarily small positive values; there is no other restriction. 

For completeness, we note that although not directly observed  in $4D$, the pressure along the extra dimension $p_{5}$, can be obtained from the equation of state (\ref{eq. of state for the core}) as
\begin{equation}
p_{5} = \frac{1}{2}(3p - \rho).
\end{equation}
Then from (\ref{int. pressure}) we find
\begin{equation}
\label{p5 for the Schw. sol}
p_{5} = \rho_{0} \frac{[(3\gamma - 1) - (5 + 9\gamma)Cr^2]}{2[1 + (2 + 3\gamma)Cr^2]}.
\end{equation}

\subsection{Boundary conditions for the Schwarzschild interior}
In terms of the  dimensionless quantities $\zeta$, $x$ and $Y$ defined  as  
\begin{equation}
\label{dimensionless quantities}
\zeta = C/a^2,\;\;\; x = ar_{b},\;\;\;Y = aR_{b}, 
\end{equation}
we have  $Cr_{b}^2 = \zeta x^2$ and
\begin{equation}
\label{N Y from the geometry}
Y = \frac{1}{x}\frac{(x + 1)^{[\varepsilon + 1 + \sigma(N - 1)]}}{(x - 1)^{[\varepsilon - 1 + \sigma(N - 1)]}}. 
\end{equation}
With this notation the continuity of the first fundamental form gives 
\begin{equation}
\label{eval. of A}
A = \frac{(1 + \zeta x^2)}{[1 + (2 + 3\gamma)\zeta x^2]}\left[\frac{x^2 - 1}{xY}\right]^{\varepsilon/[\varepsilon + \sigma(N - 1)]},
\end{equation}
and 
\begin{equation}
\label{N definition of Y}
Y = \frac{\sqrt{3 \zeta} x}{\sqrt{2 \pi(\rho_{0}/a^2)}(1 + \zeta x^2)}. 
\end{equation}
From the continuity of the second fundamental form we get
\begin{equation}
\label{Num}
\frac{(3\gamma + 1)\zeta x}{(1 + \zeta x^2)[1 + (2 + 3\gamma)\zeta x^2]} = \frac{ \varepsilon}{(x^2 - 1)},
\end{equation}
and  
\begin{equation}
\label{eq. for x}
\frac{\zeta x^2}{1 + \zeta x^2} = \frac{[\varepsilon + \sigma(N - 1)]x - 1}{(x^2 - 1)}.
\end{equation}
These equations require $x > 1$ and $[\varepsilon + \sigma(N - 1)]x \neq 1$, otherwise $C = 0$, for all values of $\varepsilon$ and $N$.

\subsection{Solution of the boundary conditions}

Dividing (\ref{Num}) by (\ref{eq. for x}) we obtain 
\begin{equation}
\label{isol Z}
\frac{3\gamma + 1}{[1 +(2 + 3\gamma)(\zeta x^2)]} = \frac{\varepsilon x}{[\varepsilon + \sigma(N - 1)]x - 1},
\end{equation} 
from which we get
\begin{equation}
\label{Z1}
(\zeta x^2) = \frac{[(N - 1)(3\gamma + 1)\sigma + 3\varepsilon \gamma]\; x - (1 + 3\gamma)}{\varepsilon x (2 + 3\gamma)}
\end{equation}
On the other hand,  from (\ref{eq. for x}) 
\begin{equation}
\label{equation for Z}
(\zeta x^2) = \frac{[\varepsilon + \sigma(N - 1)]\;x - 1}{x[x - \varepsilon - \sigma(N - 1)]}.
\end{equation}
We note that   $x \neq \varepsilon + \sigma(N - 1)$, otherwise for $\varepsilon = 1$ one would have $x = 1$. 

Equating (\ref{Z1}) and (\ref{equation for Z}) we get the desired equation for $x$. Namely,
\begin{equation}
\label{general equation for x}
[(1 + 3\gamma)(N - 1)\sigma + 3\varepsilon \gamma]\;x^2 - (1 + 3\gamma)[(N - 1)^2\sigma^2 + 3(N - 1)\varepsilon \sigma + 2\varepsilon^2 + 1]\;x + [(1 + 3\gamma)(N - 1)\sigma + 3\varepsilon(1 + 2\gamma)] = 0.
\end{equation}

\paragraph{Physical conditions:}The solutions of the above equations have to satisfy physical requirements. Positivity of mass requires $\varepsilon > 0$; positivity of $\rho_{0}$ requires $C > 0$, which translates into $\zeta > 0 $. Consequently, from (\ref{Z1}) we get
\begin{equation}
\label{positivity of density inside}
\rho_{0} > 0 \;\Rightarrow \;{\cal C}1 := [(N - 1)(3\gamma + 1)\sigma + 3\varepsilon \gamma] > 0, 
\end{equation} 
which also implies that the positivity of the coefficient in front of $x^2$.
On the other hand, from (\ref{General matter density}) it follows that 
\begin{equation}
\label{positivity of density outside}
\rho \geq 0 \;\Rightarrow \;{\cal C}2 := [3N(2 - N)\sigma^2 + 4\varepsilon(1 - N)\sigma] \geq 0.
\end{equation}

\subsubsection{Solution for $\varepsilon = 1$}
In order to facilitate the discussion, we first discuss the solution of the boundary conditions  for the Schwarzschild exterior metric.  This will allow us to compare and contrast the parameters of our model star obtained under different conditions.

In this case $\sigma = 0$, and the above equation for $x$ reduces to 
\begin{equation}
\gamma x^2  - (1 + 3\gamma)x + (1 + 2\gamma) = 0,
\end{equation}
whose solution is
\begin{equation}
\label{x for varepsilon = 1}
x = \frac{1 + 2\gamma}{\gamma}.
\end{equation}
The other ``solution" $x = 1$ is excluded. Substituting this expression into (\ref{equation for Z}) we find 
\begin{equation}
\label{Z for varepsilon = 1}
(\zeta x^2) = \frac{\gamma}{1 + 2 \gamma},
\end{equation}
and from  (\ref{N Y from the geometry}) we obtain 
\begin{equation}
\label{Y for varepsilon = 1}
Y = \frac{(3\gamma + 1)^2}{\gamma(1 + 2\gamma)}.
\end{equation}
Now, the original constants can be obtained 
from (\ref{dimensionless quantities}) and (\ref{eval. of A}). Namely,  
\begin{equation}
\label{C, rb, A}
A = \frac{1}{1 + 3\gamma}, \;\;\; \frac{3 C}{2 \pi \rho_{0}} = \left(\frac{1 + 3 \gamma}{1 + 2 \gamma}\right)^6, \;\;\;Cr^2 = \frac{\gamma}{1 + 2\gamma}\left(\frac{r}{r_{b}}\right)^2.
\end{equation}
In addition, we find 
\begin{equation}
\label{M}
r_{b} = \frac{M(1 + 2\gamma)}{2\gamma}, \;\;\;\sqrt{2\pi\rho_{0}}M = \frac{2\sqrt{3}\gamma^{3/2}(1 + 2\gamma)^{3/2}}{(1 + 3\gamma)^3},
\end{equation}
where the total mass has been obtained from (\ref{N definition of Y}). 
Thus, the boundary conditions allow us to obtain the value of all the constants in the solution in terms of the equation of the state at the center. 

\paragraph{Buchdahl limit:}

In the present case, from  (\ref{Total grav. mass}) it follows that $a = 2/M$. Consequently,  using  (\ref{Y for varepsilon = 1}) we obtain
\begin{equation}
\frac{M}{R_{b}} = \frac{2\gamma(1 + 2\gamma)}{(3\gamma + 1)^2}.
\end{equation}
In the limit $\gamma \rightarrow \infty $ we get the famous absolute upper limit 
\begin{equation}
\label{Buchdahl limit}
\left(\frac{M}{R_{b}}\right)_{max} = \frac{4}{9} \approx 0.444,
\end{equation}
discovered by Buchdahl \cite{Buchdahl}, for all static fluid spheres whose (i) energy density does not increase outward and (ii) the material of the sphere is locally isotropic.

\paragraph{Newtonian limit:} In the limit $\gamma \rightarrow 0$, from (\ref{C, rb, A}) we see  that $A \rightarrow 1$ and $(3C/2\pi \rho_{0}) \rightarrow 1$. Also, since $0 \leq (r/r_{b}) \leq 1$, it follows that $Cr \rightarrow 0$. Consequently, $e^{\nu} \rightarrow 1$ and $e^{\lambda} \rightarrow 1$. Therefore, in this limit (\ref{generalized TOV equation}) reduces to 

\begin{equation}
\label{TOV equation in Newtonian limit}
\frac{\rho_{0}M}{r^2} = - \frac{dp}{dr},
\end{equation}
which is the usual balance equation in Newtonian physics.

\subsubsection{General solution: $\varepsilon \neq 1$}
The general solution of (\ref{general equation for x}) can be written as 
\begin{equation}
\label{general solution}
x = \frac{[(N - 1)^2\sigma^2  + 3\varepsilon \sigma(N - 1) + (1 + 2\varepsilon^2)] + \sqrt{F}}{2(N - 1)\sigma + 6\gamma \varepsilon/(1 + 3\gamma)},
\end{equation}
where
\begin{equation}
\label{F}
F = (N - 1)^4\sigma^4 + 6(N - 1)^3 \varepsilon \sigma^3 + (N - 1)^2 (13\varepsilon^2 - 2)\sigma^2 + 6(N - 1)(2\varepsilon^2 - 1)\varepsilon \sigma + (2\varepsilon^2 - 1)^2 + \frac{4(3\gamma + 2)\varepsilon^2}{(3\gamma + 1)^2}.
\end{equation} 
The sign of the root in (\ref{general solution}) has been chosen positive to make sure that in the limit $\varepsilon \rightarrow 1$ we recover the solution discussed in the preceding subsection. 

Thus, for any given $(N, \varepsilon, \gamma)$ we get $x$ from (\ref{general solution}). Substituting the result into (\ref{eq. for x}) we find $(\zeta x^2)$. Using them in (\ref{dimensionless quantities}) we obtain the constants $C$, $r_{b}$ and $A$. Equation (\ref{C, rb, A}) epitomizes the fact that the first two constants depend on the total mass. In the present case  
the total mass of the configuration, i.e., the mass of the fluid sphere added to the mass of the external cloud, is obtained from (\ref{Total grav. mass}) and (\ref{N definition of Y}). The result is 
\begin{equation}
\label{general expression for the total mass}
\sqrt{2 \pi \rho_{0}}M = \frac{2\sqrt{[3 (\zeta x^2)]}\;\; \varepsilon \;  x (x - 1)^{[(\varepsilon + \sigma(N - 1) - 1]}}{(1 + \zeta x^2)(x + 1)^{[(\varepsilon + \sigma(N - 1) + 1]}}.
\end{equation}
 
\paragraph{Warning $M \neq \varepsilon  M_{Schw}$:} At this point, in order to avoid a misunderstanding, the following comment is in order. From (\ref{Total grav. mass}), one could conclude  that, since for $\varepsilon = 1$  (which corresponds to the Schwarzschild exterior metric) the  total mass is $M_{Schw} = 2/a$, then the total mass for any $\varepsilon \neq 1$ is just $M = \varepsilon M_{Schw}$. However, this is not true; it is easy to see that (\ref{general expression for the total mass})$\neq \varepsilon \times$ (\ref{M}). This is because $a$ is not a {\it fixed} constant but it is given by the boundary condition (\ref{N definition of Y}).  What is true is that for $\varepsilon = 1$, and the corresponding values of $x$ and $(\zeta x^2)$ given by (\ref{x for varepsilon = 1}) and (\ref{Z for varepsilon = 1}), we recover the mass in the  Schwarzschild
case.  

\medskip

Coming back to our discussion we calculate the mass of the star, say $M_{b}$. By virtue of the continuity of the second fundamental form, it can be easily obtained from (\ref{grav. mass}). Namely,
\begin{equation}
\label{Mb 1}
M_{b} = M\left(\frac{x - 1}{x + 1}\right)^{\sigma(1 - N)}.
\end{equation}
We note that $M_{b} = M$ only in two cases: for the Schwarzschild exterior metric $(\sigma = 0)$,  and for $N = 1$. 
 
Now, the gravitational potential at the surface of the star is given by
\begin{equation}
\label{surface gra. potential, general expression}
\frac{M_{b}}{R_{b}} = \frac{2\varepsilon x (x - 1)^{(\varepsilon - 1)}}{(x + 1)^{(\varepsilon + 1)}}.
\end{equation}

For practical reasons, it is useful to express the mass of the star in terms of the solar mass $M_{\odot}$, and its radius in kilometers (km).  They are given by
\begin{eqnarray}
\label{mass and radius in terms of solar mass and km}
M_{b} &=& \frac{1.089 \times 10^9\; \sqrt{(\zeta x^2)}\;\; \varepsilon \;  x (x - 1)^{(\varepsilon -1)}}{\sqrt{[\rho_{0}]}(x + 1)^{(\varepsilon + 1)}[1 + (\zeta x^2)]} \;\times M_{\odot}\;,\nonumber \\
R_{b} &=& \frac{0.803 \times 10^9\sqrt{(\zeta x^2)}}{\sqrt{[\rho_{0}]}[1 + (\zeta x^2)]} \;\;\;\mbox{[km]},
\end{eqnarray}
where $[\rho_{0}]$  denotes the {\it numerical} value of the energy density in $(gr/cm^3)$.

For $\varepsilon = 1$ the star is surrounded by empty space with $\rho = p_{r} = p_{\perp} = 0$ for $r > r_{b}$. However, for any $\varepsilon \neq 1$ it does have an atmosphere,  whose matter density  decreases as $\rho \sim 1/a^2r^4$ from its maximum value, $\rho_{|\Sigma^{+}}$, which is just outside $\Sigma: r = r_{b}$, to zero at infinity. For the case under consideration, from (\ref{N definition of Y}),  
(\ref{General matter density}) and using (\ref{equation for Z}) we find

\begin{equation}
\label{maximum exterior density}
\frac{\rho_{|\Sigma^{+}}}{\rho_{0}} = \frac{x \sigma^2 [3N(2 - N) + 4\varepsilon (1 - N)/\sigma]}{12\;[x - \varepsilon - \sigma (N - 1)][x(\varepsilon + \sigma (N - 1)) - 1]}.
\end{equation}
In view of the complexity of the general solution, we will analyze it for different values of $\varepsilon$ and $N$ separately.

\subsection{Factorization with $N = 1$}
In this case the above expressions are specially simple. 
As in the Schwarzschild case, we find that for any fixed $\varepsilon \neq 1$ the radius  and the mass of the fluid sphere are very sensitive to the equation of state $\gamma$. This is because $x$ decreases very fast, as $x \sim 1/\gamma$, for small $\gamma$, to a constant value for relatively large values of $\gamma$.

\subsubsection{Limiting configurations}

Let us first examine the solution of (\ref{general solution}) in the  limit $\gamma \rightarrow \infty$. 
In this case from (\ref{F}) we get $F = (2\varepsilon^2 - 1)^2$. Since we are looking for solutions that  are continuously connected to the Schwarzschild one,  we take $\varepsilon^2 > 1/2$. With this choice, from (\ref{general solution}) we obtain
\begin{equation}
\label{lim conf. for N = 1}
x = 2\varepsilon. 
\end{equation}
Then from (\ref{equation for Z}) we find 
\begin{equation}
\label{Z lim conf. for N = 1}
(\zeta x^2) = \frac{2\varepsilon^2 -1}{2\varepsilon^2}. 
\end{equation}
For $\varepsilon = 1$ we recover $x = 2$ and $(\zeta x^2) = 1/2$, which are  the corresponding limiting values in the Schwarzschild case.

Using (\ref{lim conf. for N = 1}) in (\ref{dimensionless quantities})-(\ref{N definition of Y}) we find the constants 
\begin{equation}
\label{limitting values}
C = \frac{2\varepsilon^2 -1}{2\varepsilon^2 M^2}, \;\;A \rightarrow  \frac{(2\varepsilon - 1)^{(\varepsilon + 1)}}{3\gamma (2 \varepsilon ^2 - 1)(2\varepsilon + 1)^{(\varepsilon - 1)}},
\end{equation}
as well as the mass $M$, as a function of $\varepsilon$. Namely, 
\begin{equation}
M = \frac{4 \varepsilon^3\;\sqrt{[3(2\varepsilon^2 - 1)/\pi]} \;(2\varepsilon - 1)^{(\varepsilon -2)}  }{\sqrt{\rho_{0}}(2\varepsilon + 1)^{(\varepsilon + 2)}}.
\end{equation}
From (\ref{N Y from the geometry}) and (\ref{grav. mass for N = 1}) we get the upper limit of the surface gravitational potential 
\begin{equation}
\left(\frac{M}{R_{b}}\right)_{max} = 4\varepsilon^2\frac{ (2\varepsilon - 1)^{(\varepsilon -1)}}{(2\varepsilon + 1)^{(\varepsilon + 1)}}.
\end{equation}
\paragraph{Buchdahl limit for different $\varepsilon$: }For $\varepsilon = (1.00; \;\;0.80; \;\;0.75)$ we find
\begin{eqnarray}
\sqrt{\rho_{0}}M &=& (0.145; \;\;0.135, \;\;0.112),\nonumber \\
\sqrt{\rho_{0}}R_{b} &=& (0.326, \;\;0.265; \;\;0.207),\nonumber \\
\left(\frac{M}{R_{b}}\right) &=& (\;\;4/9; \;\;0.508;  \;\;0.538),
\end{eqnarray}
respectively. Thus, with the decrease of $\varepsilon$ both the mass and the radius of the star decrease, in such a way that its surface gravitational potential increases. We note that $M/R_{b} > 1/2$, for certain values of  $\varepsilon$.  

 The red-shift $Z(r)$ of the light emitted from a point $r$ in the sphere to infinity is given by
\begin{equation}
\label{res-shift}
Z(r) = 1/\sqrt{g_{00}(r)} - 1. 
\end{equation} 
In the present case, the  red-shift of the light emitted from the boundary surface is given by
\begin{equation}
\label{Zmax}
Z _{b}= \left(\frac{2\varepsilon + 1}{2\varepsilon - 1}\right)^{\varepsilon} - 1.
\end{equation} 
Since, in the present limit\footnote{In order to avoid misunderstanding, it should be noted that for finite values of $\gamma$, like $\gamma = 1/3$ or $\gamma = 1$, the parameter $\varepsilon$ is {\it not } restricted by this condition.}, $\varepsilon^2 > 1/2$ it follows that the maximum surface red-shift is $Z_{b}  = 2.478$. This indicates that, although the surface gravitational potential is $M/R_{b} > 1/2$ (instead of the usual $M/R_{b} < 1/2$), the star is {\it not} a black hole.

\paragraph{Newtonian limit for different $\varepsilon$:}
For $\gamma \rightarrow 0$, from (\ref{general solution}) with $N = 1$, 
 we get
\begin{equation}
x \rightarrow \frac{2\varepsilon^2 + 1}{3 \varepsilon \gamma}. 
\end{equation}
Substituting this into (\ref{Z1}), or (\ref{equation for Z}), we find
\begin{equation}
(\zeta x^2) \rightarrow \frac{3\varepsilon^2 \gamma}{2\epsilon^2 + 1}.
\end{equation}
For $\varepsilon = 1$ these quantities become $x \rightarrow 1/\gamma$, $(\zeta x^2) \rightarrow \gamma$, which are  identical to their Newtonian values in the Schwarzschild case. In this limit,  (\ref{N Y from the geometry}) yields $Y \sim x$  
Substituting these expressions into (\ref{eval. of A}) and (\ref{N definition of Y}) we get
\begin{equation}
A \rightarrow 1, \;\;\;\frac{3C}{2\pi \rho_{0}} \rightarrow 1,\;\;\;Cr^2 \rightarrow \frac{3\varepsilon \gamma}{2 \varepsilon^2 + 1}\left(\frac{r}{r_{b}}\right)^2
\end{equation}
Therefore, for $N = 1$ in the limit $\gamma \rightarrow 0$ we recover flat spacetime as well as the Newtonian balance equation (\ref{TOV equation in Newtonian limit}). 

\subsubsection{Effects of $\varepsilon \neq 1$ on neutron star parameters}

The second important question in this work is how going away from $\varepsilon = 1$ may affect the star parameters. In order to study this question, we have calculated the surface gravitational potential $M/R_{b}$, the red-shifts at the center $Z_{c}$ and at the boundary $Z_{b}$ for different values of $\gamma$. The mass and radius of the fluid sphere are given by 
(\ref{mass and radius in terms of solar mass and km}). To
 obtain some numerical values for $M$ and $R_{b}$ we have taken $\rho = 2 \times 10^{14} gr/cm^3$ which has been used in different models of neutron stars \cite{Durgapal}, \cite{New analytic}. Table $1$ shows the results for the Schwarzschild solution $(\varepsilon = 1)$, which we use to compare and contrast with the results of Tables $2$ and $3$ for $\varepsilon = 0.9$ and $\varepsilon = 0.8$, respectively. 

\begin{center}
\begin{tabular}{|c|c|c|c|c|c|} \hline
\multicolumn{6}{|c|}{\bf Table 1: Star parameters for the Schwarzschild exterior} \\ \hline
 \multicolumn {1}{|c|}{$\gamma = p_{c}/\rho_{c}$} & 
$M/R_{b}$ & $ Z_{c}$ & $Z_{b}$ &$R_{b}(km)$&
 \multicolumn{1}{|c|}{$M_{g}(M_{\odot})$} \\ \hline\hline
                    $1/10$ & $0.14$& $0.30$ & $0.18$& $15.13$&$1.46$   \\ \hline
$1/3$ &$0.28$ & $1$   & $0.50$& $21.16$&$3.99$   \\ \hline
 $1 $ &$0.38$ & $3.00$ & $1.00$ &$24.59$&$6.25$ \\ \hline
$3 $ &$0.42$ & $9.00$ & $1.50$ &$26.02$&$7.41$ \\ \hline
$10 $ &$0.44$ & $30$ & $1.82$ &$26.54$&$7.87$ \\ \hline

$\infty $ &$4/9$ & $\infty$ & $2.00$ &$26.77$&$8.08$ \\ \hline
\end{tabular}
\end{center}
Table $1$ shows how the physical features of a neutron star, with a vacuum  Schwarzschild exterior, depend on the equation of state at the center.  

\begin{center}
\begin{tabular}{|c|c|c|c|c|c|c|} \hline
\multicolumn{7}{|c|}{\bf Table 2: Star parameters for $\epsilon =0.9$ and $N = 1$}\\ \hline
 \multicolumn {1}{|c|}{$\gamma = p_{c}/\rho_{c}$} & 
$M/R_{b}$ & $ Z_{c}$ & $Z_{b}$ &$R_{b}(km)$ & $M_{b}(M_{\odot})$&
 \multicolumn{1}{|c|}{$\rho_{|\Sigma^{+}}/\rho_{0}$} \\ \hline\hline
                    $1/10$ & $0.13$& $0.28$ & $0.17$& $14.56$&$1.32$&$0.007$   \\ \hline
$1/3$ &$0.28$ & $0.91$   & $0.47$& $20.34$&$3.71$&$0.024$   \\ \hline

 $1 $ &$0.37$ & $2.68$ & $0.95$ &$23.57$&$5.99$&$0.061$ \\ \hline

$3 $ &$0.43$ & $7.88$ & $1.48$ &$24.85$&$7.27$&$0.117$ \\ \hline

$10 $ &$0.46$ & $25.88$ & $1.85$ &$25.26$&$7.82$&$0.167$ \\ \hline
$\infty $ &$0.47$ & $\infty$ & $2.09$ &$25.40$&$8.07$&$0.204$ \\ \hline
\end{tabular}
\end{center}

\begin{center}
\begin{tabular}{|c|c|c|c|c|c|c|} \hline
\multicolumn{7}{|c|}{\bf Table 3: Star parameters for $\epsilon = 0.8$ and $N = 1$}\\ \hline
 \multicolumn {1}{|c|}{$\gamma = p_{c}/\rho_{c}$} & 
$M/R_{b}$ & $ Z_{c}$ & $Z_{b}$ &$R_{b}(km)$ & $M_{b}(M_{\odot})$&
 \multicolumn{1}{|c|}{$\rho_{|\Sigma^{+}}/\rho_{0}$} \\ \hline\hline
                    $1/10$ & $0.13$& $0.28$ & $0.17$& $14.56$&$1.32$&$0.007$   \\ \hline
$1/3$ &$0.26$ & $0.80$   & $0.42$& $19.25$&$3.34$&$0.054$   \\ \hline

 $1 $ &$0.37$ & $2.26$ & $0.87$ &$22.04$&$5.54$&$0.149$ \\ \hline

$3 $ &$0.44$ & $6.22$ & $1.41$ &$22.71$&$6.82$&$0.333$ \\ \hline

$10 $ &$0.48$ & $19.62$ & $1.86$ &$22.71$&$7.34$&$0.576$ \\ \hline
$\infty $ &$0.51$ & $\infty$ & $2.23$ &$21.79$&$7.50$&$0.857$ \\ \hline
\end{tabular}
\end{center}
Tables $2$ and $3$ contain the discontinuity of the energy density at the boundary calculated from (\ref{maximum exterior density}). We see that as the surface gravitational potential increases, $\rho_{|\Sigma^{+}}$  also increases, that is the atmosphere  gets more concentrated near the surface, which is physically reasonable.

\subsection{Factorization with $N \neq 1$ and $\varepsilon \neq 1$}

The energy condition (\ref{positivity of density inside}) requires the denominator in  (\ref{general solution}) to be positive. However, it vanishes at $\gamma = \bar{\gamma}$ given by
\begin{equation}
\bar{\gamma} = \frac{(1 - N)\sigma}{3[\varepsilon + (1 - N)\sigma]}.
\end{equation}
We note that $\bar{\gamma} = 0$ only for $N = 1$ and the Schwarzschild exterior. 
Since $\sigma$ is positive for $N < 1$, and negative for $N > 1$, it follows that $(1 - N)\sigma > 0$. Consequently, for any other $N$ and $\varepsilon$  we have $\bar{\gamma} > 0$. 

On the other hand, since $x$ is positive for $\gamma > \bar{\gamma}$ and negative for $\gamma < \bar{\gamma}$, it follows that $\gamma$ has a positive lower bound, i.e.,
\begin{equation}
\label{lower bound for gamma}
\gamma > \bar{\gamma}.
\end{equation}
Thus, in accordance with our discussion in section $3$, we conclude that the factorization with $N \neq 1$, and $\varepsilon \neq 1$, is incompatible with the Newtonian limit, which requires $\gamma \rightarrow 0$. Let us consider some particular cases with more detail.

\subsubsection{Factorization with $N = 0$}

 This corresponds to the case where the metric of the physical spacetime is identified with the metric induced in $4D$, which  is the typical approach in induced-matter and braneworld theories. In this case the exterior metric 
(\ref{eff. spacetime for Davidson and Owen solution}) can be written as
\begin{equation}
\label{exterior metric N = 0}
ds^2 = \left(\frac{ar - 1}{ar + 1}\right)^{2\varepsilon} dt^2 - \frac{(ar + 1)^2(ar - 1)^2}{a^4 r^4}\left(\frac{ar + 1}{ar - 1}\right)^{[\varepsilon + \sqrt{4 - 3\varepsilon^2}]}[dr^2 + r^2d\Omega^2],
\end{equation}
where we have used that, from (\ref{sigma in terms of varepsilon}), in the present case 
\begin{equation}
\label{sigma for N = 0}
\sigma = \frac{\varepsilon - \sqrt{4 - 3\varepsilon^2}}{2},
\end{equation}
which requires 
$\varepsilon  \leq 2/\sqrt{3}$. 

Now, from (\ref{positivity of density inside}) and (\ref{positivity of density outside}) we get the constraints 
\begin{eqnarray}
\label{positivity of density for N = 0}
{\cal C}1 &:=& - (3\gamma + 1)\sigma + 3\varepsilon \gamma > 0, \nonumber \\
{\cal C}2 &:=& 4 \varepsilon \sigma \geq 0.
\end{eqnarray}
From which we find that the positivity of the energy density inside of the fluid sphere requires
\begin{equation}
\gamma > \bar{\gamma} = \frac{\varepsilon - \sqrt{4 - 3\varepsilon^2}}{3(\varepsilon + \sqrt{4 - 3\varepsilon^2})}.
\end{equation}
We note that $x$ is positive for $\gamma > \bar{\gamma}$ and negative for $\gamma < \bar{\gamma}$. In order to get a better understanding of these constraints, let us take  $\varepsilon = (0.8, \;\;0.9,\;\;1,\;\;1.1,\;\;1.15, \;\;2/\sqrt{3})$.  We find
\begin{eqnarray}
\gamma &>& (- 0.09, \;\;- 0.05, \;\;0, \;\;0.09, \;\;0.23, \;\;1/3),\nonumber \\ 
\sigma &=& (- 0.32, \;\;- 0.18, \;\;0, \;\;0.25, \;\;0.48, \;\;1/\sqrt{3}),
\end{eqnarray}
respectively. 
Thus, the positivity of the energy density outside of the fluid sphere requires  $\varepsilon \geq 1$. For $\varepsilon = 1$ we recover the usual Schwarzschild picture, as expected. However,  for $\varepsilon > 1$, $\gamma$ has a lower positive bound, e.g.,  $\bar{\gamma} =  0.23$ for $\varepsilon = 1.15$, which increases with the increase of $\varepsilon$. Certainly these models have no Newtonian limit.

\subsubsection {Factorization with $N = 2$}

In this case 
\begin{equation}
\label{sigma for N = 2}
\sigma = \frac{- \varepsilon + \sqrt{4 - 3\varepsilon^2}}{2}.
\end{equation}
Substituting this into (\ref{eff. spacetime for Davidson and Owen solution}) we find that the line element is  identical to (\ref{exterior metric N = 0}), which is consistent with the symmetry under the change $N \rightarrow (2 - N)$ noted in subsection $2.2$. Also, both factorizations, $N = 0$ and $N = 2$,  yield the same equation of state  (\ref{equation of state}),  and have the same lower limit on $\gamma$. 

The discussion in this section epitomizes a situation frequently found in physics. Namely, that the analysis of some particular ``simple" example can be much more involved, in technical details, than the general case.  
However, the effort is worthwhile: the main conclusions from our uniform-density model are consistent with the general results presented in section $3$. Specifically, (i) Only the factorization with $N = 1$ produces stars models that have a Newtonian limit; (ii) the factorizations with $N$ and $(2 - N)$ are physically indistinguishable.

\section{Summary and concluding remarks}

For theories in more than four-dimensions the crucial question is, how to recover the physics of our $4$-dimensional world.  The answer to this question is important in order to be able to calculate and predict specific effects from extra dimensions. For example, the possible existence of extra ``non-gravitational" forces on test particles and their effect on the classical tests of relativity and the principle of equivalence \cite{Wesson book}, \cite{JPdeL}. 

In this paper, we have considered this question from the point of view of the stellar structure. Namely, since Birkhoff's  theorem is no longer valid in more than four-dimensions, there is no an unique asymptotically flat vacuum solution with spatial spherical symmetry.  In other words, regardless of whether we choose to work in Kaluza-Klein gravity \cite{Wesson-JPdeL}, STM \cite{Wesson book}, braneworld models \cite{Cristiano}, \cite{Bruni}  or others theories with curvature corrections \cite{Kofinas}, the fact is that the {\it effective} four-dimensional theory allows the existence   
of  different possible, non-Schwarzschild, scenarios for the description of the spacetime outside of a spherical star, contrary to four-dimensional general relativity. 

In five-dimensional Kaluza-Klein gravity the well-known solutions due to Kramer-Davidson-Owen-Gross-Perry seem to represent the natural generalization of the Schwarzschild spacetime \cite{Wesson-JPdeL}-\cite{JPdeL:gr-qc/0512067}. Even in this simple case of spherical symmetry, in ordinary three space, there are various possible options leading to different exteriors for a spherical star. A popular assumption is that we recover our $4D$ spacetime by going onto a hypersurface $y$ = constant.  With this assumption, the information about the fifth dimension is consolidated in the  nonlocal stresses induced in $4D$ from the Weyl curvature in $5D$.  As a consequence,  even in the absence of matter, the exterior of a spherical star, like our sun,  is {\it not} in general an empty Schwarzschild spacetime. 

However, early in our research we found  that this simple approach leads to contradictions in the Newtonian limit. 
Therefore, in section $2$  we generalized our discussion by considering, as a possible exterior for  a static spherical star, the  family of asymptotically flat metrics represented by (\ref{eff. spacetime for Davidson and Owen solution}). These are  conformal to the induced metric (\ref{general factorization}) and constitute a generalization to a number of approaches in the literature (see for example \cite{Mashhoon}-\cite{Dolan}).

The main question under scrutiny here  was whether the existence of stars can be used to constraint the number of possible candidates for exteriors. Using the junctions  conditions at the boundary of the star, and without making any assumptions about the interior metric, we have showed  that (\ref{The only exterior metric compatible with New. Lim}) is the only exterior metric compatible with both; the weak-field Newtonian limit as well as the general relativistic Schwarzschild limit. All the other metrics with $N \neq 1$ do not have a Newtonian counterpart, although they are consistent with the Schwarzschild limit.

It should be noted that the resulting effective matter (\ref{external density for N = 1}), outside of a spherical star,   is ``similar" to Schwarzschild vacuum in the sense that it has no effect on gravitational interactions. In fact, it 
satisfies the  equation of state $\rho = -(p_{r} + 2p_{\perp})$, which as we can see from (\ref{standard gravitational mass}) corresponds to  $M = 0$. This is the generalized version of the well-known equation of state $\rho =  - 3 p$ for isotropic fluid, which has been discussed in different contexts (see for example \cite{Gott}-\cite{WessonEssay}). One can speculate that this exterior (with $N = 1$) is more realistic than the Schwarzschild one, because instead of being absolutely empty $(\rho = p_{r} = p_{\perp} = 0)$, it is consistent with the existence of quantum zero-point fields \cite{WessonEssay}.

The question may arise about the generality of our result. Namely, what would happen if we assume that the effective four-dimensional metric is like
\begin{equation}
g^{eff}_{\mu\nu} = F(\Phi)g^{ind}_{\mu\nu},
\end{equation} 
where $F$ is some ``function" of $\Phi$?. It is not difficult to show, using the boundary conditions, that the only function compatible with a  Newtonian limit is $F \sim \Phi$, i.e., the factorization with $N = 1$. 

Another important question here is what are the consequences of going away from the Schwarzschild exterior metric. Using a uniform-density model for the interior of the star,  in Tables $1$-$3$ we show  how some physical quantities as the mass, radius and surface gravitational potential of a star are affected by an extra dimension. For the same equation of state  at the center, we see that both the mass and the radius of a the neutron star are reduced as we separate from the Schwarzschild vacuum exterior. 
The most striking feature  is that, in principle, as a consequence of the extra dimension, stars may exist with gravitational potentials larger than $1/2$ without being black holes. Certainly, this can provide some clues for the observational test of the theory. 

Regarding the classical tests of general relativity and other experimental observations, within the context of Kaluza-Klein gravity \cite{Wesson book}, they have been calculated under the assumption that the physical spacetime outside an astrophysical object, like our sun, was described by the spacetime part of (\ref{Davidson and Owen solution}). Our results here suggest to consider the metric  (\ref{The only exterior metric compatible with New. Lim})
\[
ds^2 = \left(\frac{ar - 1}{ar  + 1}\right)^{2\varepsilon}dt^2  - \frac{1}{a^4r^4}\frac{(ar + 1)^{2[\varepsilon + 1]}}{(ar - 1)^{2[\varepsilon - 1]}}[dr^2 + r^2d\Omega^2].
\]
As far as we know this line element has never been used in this kind of discussions.

\renewcommand{\theequation}{A-\arabic{equation}}
  \setcounter{equation}{0}  
  \section*{Appendix A: Possible non-vacuum exteriors of a spherical star.}  

 An observer in $4D$, who is not directly aware of the existence of an extra dimension, will interpret the metric (\ref{eff. spacetime for Davidson and Owen solution}) as if it were governed 
 by an effective energy-momentum tensor.

\begin{equation}
\label{General matter density}
8\pi T^{0}_{0} = \frac{a^6 r^4\sigma^2[3N(2 - N) + 4\varepsilon (1 - N)/\sigma]}{(ar + 1)^4(ar - 1)^4}\left(\frac{ar - 1}{ar + 1}\right)^{2[\varepsilon + \sigma (N - 1)]},
\end{equation}

\begin{equation}
\label{general radial pressure}
8 \pi T_{1}^{1} = - \frac{4 a^5 r^3\sigma^2 \left\{(1 -N)[ar(4\sigma - 2\varepsilon - N\sigma)/(2\sigma) + (a^2r^2 +1)/\sigma] + (3/4)N^2 ar\right\}}{(ar + 1)^4(ar - 1)^4} \left(\frac{ar - 1}{ar + 1}\right)^{2[\varepsilon + \sigma (N - 1)]},
\end{equation}

\begin{equation}
\label{general tangential  pressure}
8 \pi T_{2}^{2} =  - \frac{2 a^5r^3 \sigma^2\left\{2ar[(2\varepsilon + N\sigma)(1 - N)/(2\sigma) - 1] - (1 - N)(a^2r^2 + 1)/\sigma) + (1/2)N^2 ar\right\}}{(ar + 1)^4(ar - 1)^4}\left(\frac{ar - 1}{ar + 1}\right)^{2[\varepsilon + \sigma (N - 1)]},
\end{equation}
and $T_{3}^{3} = T_{2}^{2}$,  as expected by virtue of the spherical symmetry.  These equations show that  $T_{2}^{2} \neq  T_{1}^{1}$ for {\it any} value of $N$ and $\varepsilon \neq 1$. Thus, the effective matter  behaves  as an anisotropic fluid,  which can be described by an effective energy-momentum tensor of the form
\begin{equation}
\label{EMT for anisotropic matter}
T_{\mu\nu} = (\rho + p_{\perp})u_{\mu}u_{\nu} - p_{\perp}g_{\mu\nu} + (p_{r} - p_{\perp})\chi_{\mu}\chi_{\nu}, 
\end{equation}
where $u^{\mu}$ is the four-velocity; $\chi^{\mu}$ is a unit spacelike vector orthogonal to $u^{\mu}$; $\rho$ is the energy density; $p_{r}$ is the pressure in the direction of $\chi_{\mu}$, and $p_{\perp}$ is the pressure on the two-space orthogonal to $\chi_{\mu}$.  Since $T_{01} = 0$, it follows that the fluid is at rest in then system of reference of (\ref{eff. spacetime for Davidson and Owen solution}). Therefore, we can always choose $u^{\mu} = \delta^{\mu}_{0}/\sqrt{g_{00}}$ and $\chi^{\mu} = \delta^{\mu}_{1}/\sqrt{(- g_{11})}$. Thus, 
\begin{equation}
\label{components of EMT}
T^0_{0} = \rho,\;\;\; T^1_{1} = - p_{r}, \;\;\; \mbox{and} \;\;\;T^2_{2} = T^3_{3} = - p_{\perp}. 
\end{equation}

\paragraph{Equation of state:} 

Now, from (\ref{General matter density})-(\ref{general tangential pressure}), we find the equation of state of the effective matter. It is  
\begin{equation}
\label{equation of state}
\rho = f(N, \sigma)(p_{r} + 2p_{\perp}),
\end{equation}
with
\begin{equation}
\label{def. of f(N, k)}
f(N, \sigma) = \frac{4\varepsilon (1 - N) - 3N\sigma(N - 2)}{4\varepsilon(1 - N) + 3 N \sigma(N - 2)},
\end{equation}
where $\sigma$ is given by (\ref{sigma in terms of varepsilon}). 
Note that $f = 1$ for $N = 0$ and  $N = 2$ the equation of state becomes
\begin{equation}
\label{equation of state for N = 0}
\rho = p_{r} + 2p_{\perp}.
\end{equation}
For $N = 1$ we get 
\begin{equation}
\label{Eq. of state for N = 1}
\rho = - (p_{r} + 2p_{\perp}).
\end{equation}
In general, for any value of $N$ the equation of state approximates (\ref{equation of state for N = 0}), for ``small" values of $\sigma$, i.e., for $\varepsilon \approx 1$.

\paragraph{Gravitational mass:} 

In $4D$,  the gravitational mass inside a $3D$ volume $V_{3}$ is given by the Tolman-Whittaker formula viz., 

\begin{equation}
\label{standard gravitational mass}
M(r) = \int{(T^0_0 - T^1_1 - T^2_2 - T^3_3)}\sqrt{- g_{4}}dV_{3}.
\end{equation}
In isotropic coordinates, with $g_{00} = e^{\nu(r)}$ and  $g_{11} = - e^{\lambda(r)}$,  this simplifies to 
\begin{equation}
\label{grav. mass in terms of the metric}
M(r) = \frac{1}{2}r^2 e^{(\nu  + \lambda)/2}\nu',
\end{equation}
 where the prime $'$ denotes derivative with respect to $r$. Using this equation, for the metric (\ref{eff. spacetime for Davidson and Owen solution})  we obtain 

\begin{equation}
\label{grav. mass}
M(r) =  \frac{2\varepsilon}{a}\left(\frac{ar - 1}{ar +1}\right)^{\sigma(1 - N)}.
\end{equation}
Here, we require $\varepsilon > 0$ in order to ensure the positivity of $M$.

\end{document}